\documentclass{article}
\begin{document}

\newcommand{\Author}    {\large\bf Sava\c{s} Arapo\~glu$^{a}$ and Cihan Sa\c cl\i
o\~glu$^{a,b}$}

\def\toprule{\noalign{\hrule \medskip}}
\def\midrule{\noalign{\medskip\hrule }}
\def\botrule{\noalign{\medskip\hrule }}
\setlength{\parskip}{\medskipamount}
\newcommand{\ud}{{\mathrm{d}}}


\begin{titlepage}
%
%
%
%
\begin{center}
    \huge\bf\boldmath
A Ramond-Neveu-Schwarz string with one end fixed
\end{center}
\bigskip
%
%
\begin{center}
\Author
\end{center}

\begin{center}
{\small
$^a$ Physics Department, Bo\~gazi\c ci University, 34342 Bebek,
Istanbul, Turkey\\
$^b$ Feza G\"ursey Institute, TUBITAK--Bo\~gazi\c ci University,
81220,
Istanbul, Turkey }

\end{center}

\bigskip
%
%
\begin{abstract}

 We study an RNS string with one end fixed on a $D0$-brane and
the other end free as a qualitative guide to the spectrum of
hadrons containing one very heavy quark. The mixed boundary
conditions break half of the world-sheet supersymmetry.
Boson-fermion masses can still be matched if space-time is 9
dimensional; thus $SO(8)$ triality still plays a role in the
spectrum, although full space-time supersymmetry does not survive.
We quantize the system in a temporal-like gauge where $X^0 \sim
\tau$.  Only odd $\alpha$ and even $d$ R modes remain, while the
NS oscillators $b$ become odd-integer moded.  Although the gauge
choice eliminates negative-norm states at the outset, there are
still even-moded Virasoro and even(odd) moded super-Virasoro
constraints to be imposed in the NS(R) sectors. The Casimir energy
is now positive in both sectors; there are no
tachyons. States for $\alpha' M^2 \leq 3$ are explicitly
constructed and found to be organized into $SO(8)$ irreps by
(super)constraints, which include a novel ``$\sqrt{L_0}$" operator
in the NS and $\Gamma^0 \pm I$ in the R sectors. GSO projections
are not allowed.  The pre-constraint states above the ground state have 
matching multiplicities, indicating spacetime supersymmetry is broken by 
the (super)constraints.  A distinctive physical signature 
of the system is a slope twice that of the open RNS string. When
both ends are fixed, all leading and subleading trajectories are
eliminated, resulting in a spectrum qualitatively similar
to the $J/\psi$ and $\Upsilon$ particles.

\end{abstract}

\end{titlepage}

\section{Introduction}

Bosonic String theory was originally discovered  \cite{Nambu,
Susskind} while trying to account for observed properties of
hadron dynamics, and it was indeed qualitatively successful in
reproducing features of hadron physics such as linear Regge
trajectories for mesons, amplitudes with Dolen-Horn-Schmid duality
\cite{Dolen} and desired high-$s$ and $t$ behaviors and poles
\cite{Ven}. The nearly-massless up and down quarks and antiquarks
were incorporated into the string picture by being placed at open
string endpoints, where Neumann boundary conditions ensured their
moving at the speed of light.  Harari-Rosner diagrams
\cite{Har,Ros} were useful for keeping track of internal quantum
numbers and visualising duality between Regge pole exchanges and
resonances in the $s$-channel.

There was an obvious need to find an extension of the string model
that could encompass baryons, which had masses similar to those of
the mesons (except for the very light pion), and also lay
on Regge trajectories with approximately the same slope.  These
similarities suggested a new kind of symmetry (albeit partially
broken, in view of the non-vanishing mass differences) between
bosons and fermions, and string models featuring such a symmetry
were constructed by Ramond \cite{Ramond} and Neveu and Schwarz
\cite{NS}.  It gradually became clear, however, that
hadron dynamics was governed by a gauge theory, and
superstring theory was elevated to the status of a promising candidate
for describing quantum gravity and other fundamental physical phenomena at
distances of $10^{-34}$ m.  The partial success of strings in modelling
hadrons was attributed to gluons forming string-like flux tubes between
quarks.

It may be instructive to return to this oldest use
of string theory as a phenomenological guide for the study of new
generations of baryons and mesons containing at least one heavy
(of the top, bottom or charmed variety) and one light quark. In an
earlier work \cite{Gursoy} we modeled such mesons as excitations of an
open string with one end fixed on a $D0$-brane and the other end free.
This gave rise to testable predictions such as a doubled Regge slope and a 
reduction in states due to the restriction to odd oscillators.  
Whether the description is of any merit will become apparent when higher
spin excitations
of heavy quark bearing mesons are found and studied.  Strings with mixed
boundary conditions have also been considered in
\cite{Fair,Sieg,Klein,Froh}.

In this note, we examine the Ramond-Neveu-Schwarz (RNS) version of
\cite{Gursoy}.  Our main motivation and hope is that this may
provide qualitative hints about the spectrum of baryons and mesons
with a single $c,b$ or a $t$ quark, although we of course do not expect
it to provide a heavy-hadron  model that is realistic in all its details.

Apart from its possible phenomenological usefulness, the system is
also of some intrinsic string-theoretic interest in a number of respects.  One of these is that 
the system we are 
studying can be thought of as  an open RNS string viewed from a frame where its mid-point is at 
rest if one disregards one half of the string.  Exactly the same ``half a Neumann-Neumann string" 
picture with even 
modes in the R and odd modes in the NS sector is encountered in String Field Theory 
\cite{Bars}, where the interaction of two strings is effected by joining them up to the mid-point 
and leaving the other halves free. 
Another point of interest is the use of a gauge halfway
between the light-cone and the old covariant treatments.  This gauge 
naturally suggests itself in the mixed ND string.  
The old
covariant method of quantization of even the standard RNS string
is described only very sketchily in standard reference works such
as \cite{GSW,Pol,Pol2}; the detailed implementation of the constraints
peculiar to our problem  and the subsequent emergence of the physical
spectrum turns out to be surprisingly intricate.   

The plan of the paper is as follows: In section 2 we obtain the
mode expansions and the restricted SUSY transformations allowed by
our novel boundary conditions.  We then impose a gauge we might
call ``the rest-frame gauge". Section 3 is devoted to the Virasoro
and super-Virasoro constraints in the NS and R sectors.  Some of
the new features that emerge include the limitation of the
Virasoro algebra to even modes, while the modes of the R and NS
constraints can only be odd and even respectively. The
Virasoro and super-Virasoro algebras turn out to have
non-vanishing anomalies in all dimensions for reasons we will
discuss.  In section 4, we work out the low-lying members of the
NS and R spectra.  The organization and projection of physical
states into specific $SO(D-1)$ irreps via the constraints involves
novelties such as an NS constraint that may be regarded as ``a
bosonic-sector square root of $L_0$" and an R constraint which is a
truncated form of the usual Ramond-Dirac operator $F_0$.  The
spectrum does not exhibit full space-time supersymmetry, but it is
at least possible to match the masses of bosons and fermions if
$D-1=8$, which means aspects of $SO(8)$ triality are still
reflected in the states. Because of changes in the modes and their Casimir energies, both the NS
the R ground states become massive. Both the absence of purely transverse 
states 
and and a spacetime dimension less than 10 suggest are features of 
consistent string theories in sub-critical dimensions.
For higher states, we present a level multiplicity formula which counts the number of states 
before the constraints are imposed. Remarkably, the number of 
unconstrained bosonic and fermionic states (except for the bosonic 
ground state) are equal up to all the levels 
we have calculated.  This indicates  
that the partial breaking of world-sheet
supersymmetry through mixed boundary conditions is reflected in the space-time 
picture in a surprisingly indirect way through the superconstraints.   
The paper ends with concluding remarks in
section 5.

\section{The classical RNS string with one end fixed}
\subsection{Equation of motion:}

We start with standard material to set our notation and
conventions.  In the superconformal gauge the action reads
\begin{equation}
S=-\frac{1}{2\pi}\int
d^{2}\sigma\left\{\partial_{\alpha}X_{\mu}\partial^{\alpha}X^{\mu}-i\overline{\Psi}^{\mu}\rho^{\alpha}\partial_{\alpha}\Psi_{\mu}\right\},
\end{equation}
where $\rho^{\alpha}$ represents the two dimensional Dirac
matrices and a convenient basis satisfying
$\left\{\rho^{\alpha},\rho^{\beta}\right\}=-2\eta^{\alpha\beta}$
is
\begin{center}
$\rho^{0}=\left[
\begin{array}{cc}
0 & -i \\
i & 0%
\end{array}%
\right]$, $\rho^{1}=\left[
\begin{array}{cc}
0 & i \\
i & 0%
\end{array}%
\right]$.
\end{center}
We will denote the upper and lower components of $\Psi$ as
$\Psi_{-}$ and $\Psi_{+}$, respectively.  We should note that the
Dirac operator, $i\rho^{\alpha}\partial_{\alpha}$, is real in this
representation; thus, the components of the Majorana world sheet
spinor may be taken as real. In this paper, our focus will be on
these spinor degrees of freedom since the bosonic fields have
already been treated in \cite{Gursoy}.

The fermion equation of motion derived from (1) is simply the two
dimensional Dirac equation and in the chosen basis it splits up
into the following equations:
\begin{eqnarray}
\frac{ 1}{2}  \partial_{+}\Psi_{-}^{\mu}\equiv(\frac{\partial}{\partial\sigma}+\frac{\partial}{\partial\tau})\Psi_{-}^{\mu}&=&0,\\
-\frac{1}{2}
\partial_{-}\Psi_{+}^{\mu}\equiv(\frac{\partial}{\partial\sigma}-\frac{\partial}{\partial\tau})\Psi_{+}^{\mu}&=&0.
\end{eqnarray}
The equation of motion for the bosonic field in the same basis is
\begin{equation}
\partial_{+}(\partial_{-}X^{\mu})=\partial_{-}(\partial_{+}X^{\mu})=0.
\end{equation}
\subsection{Boundary conditions and mode expansions:}

The boundary conditions for the the bosonic fields are as in
\cite{Gursoy}: We locate the infinitely massive ``quark" at the
$\sigma=0$ end and identify this point with the origin of target
space coordinates through the Dirichlet boundary condition
\begin{equation}
X^{i}(0, \tau)=0,
\end{equation}
while we adopt the Neumann boundary condition
\begin{equation}
\partial_{\sigma}X^{\mu}(\sigma_{max}, \tau)=0
\end{equation}
for the massless end.  If we use the usual integrally moded
oscillators to expand $X^\mu$, these two conditions lead to the
requirement that $\alpha_{n}^{\mu}=0$ for even n and
$\sigma_{max}=\pi/2$ (another option, adopted in \cite{Fair} among others,
is to maintain the usual range $[0,\pi]$ for $\sigma$, in which case the
$\alpha_{n}^{\mu}$ oscillators become half-integrally moded; our
choice is to work with the odd-numbered subset of the original
integral modes rather than to introduce an entirely new set of
oscillators). As mentioned in \cite{Gursoy}, the disappearance of the
even modes is quite similar to the same phenomenon in the
elementary quantum mechanics problem of a half-oscillator
potential $V=\frac{1}{2} kx^2$ for $x\geq 0$, $V=\infty$ for $x<
0$. As in the usual RNS string, vanishing of the surface term
requires $\Psi_{+}^{\mu}=\pm\Psi_{-}^{\mu}$ at each end, and
without loss of generality, we set $\Psi_{+}^{\mu}(0, \tau
)=\Psi_{-}^{\mu}(0, \tau)$ for $\sigma=0$ and then consider the
two cases depending on the relative sign at $\sigma_{max}=\pi/2$.
Leaving the question of how to reconcile the choice at $\sigma=0$
with world-sheet supersymmetry to the next section, the two cases
arising from the sign choices are the following:

(i) The Ramond (R) boundary condition
\begin{equation}
\Psi_{+}^{i}(\frac{\pi}{2},\tau)=\Psi_{-}^{i}(\frac{\pi}{2},\tau)
\end{equation}
leads to the exclusively even-moded expansion
\begin{eqnarray}
\Psi_{-}^{i}(\sigma,\tau)&=&\sum_{n:even}d_{n}^{i}e^{-in(\tau-\sigma)},\\
\Psi_{+}^{i}(\sigma,\tau)&=&\sum_{n:even}d_{n}^{i}e^{-in(\tau+\sigma)}.
\end{eqnarray}

(ii) The Neveu-Schwarz (NS) boundary condition
\begin{equation}
\Psi_{+}^{i}(\frac{\pi}{2},\tau)=-\Psi_{-}^{i}(\frac{\pi}{2},\tau)
\end{equation}
results in the odd-integer modes
\begin{eqnarray}
\Psi_{-}^{i}(\sigma,\tau)&=&\sum_{r:odd}b_{r}^{i}e^{-ir(\tau-\sigma)},\\
\Psi_{+}^{i}(\sigma,\tau)&=&\sum_{r:odd}b_{r}^{i}e^{-ir(\tau+\sigma)}.
\end{eqnarray}
We note that while the boundary conditions merely restrict the
Ramond oscillators $d^{i}_{n}$ to even and the $\alpha_{n}^{i}$ to
odd modes, the Neveu-Schwarz sector actually changes from
half-integral to odd integral oscillators.  This will lead to a
massless NS and a massive R ground state, reversing the usual RNS
results.

\subsection{Broken global world-sheet SUSY:}
The action is invariant under the infinitesimal world-sheet
supersymmetry transformations
\begin{equation}
\delta X^{\mu}=\overline{\epsilon}\Psi^{\mu},
\end{equation}
\begin{equation}
\delta\Psi^{\mu}=-i\rho^{\alpha}\partial_{\alpha}X^{\mu}\epsilon.
\end{equation}
with $\epsilon$ a constant anticommuting spinor. Let us examine
the space components of (13) and (14) first. For fermionic fields
(13) combined with (5) gives
\begin{equation}
\delta X^{i}(0,\tau)_{\sigma = 0} =
(\epsilon_{+}^{\dag}\Psi_{-}^{i}-\epsilon_{-}^{\dag}
\Psi_{+}^{i})_{\sigma=0}=0.
\end{equation}
This would normally be satisfied either by

(i)
\begin{equation}
\Psi_{+}^{i}(0,\tau)=\Psi_{-}^{i}(0,\tau),
\end{equation}
and
\begin{equation}
\epsilon_{+}=\epsilon_{-},
\end{equation}
or by

(ii)
\begin{equation}
\Psi_{+}^{i}(0,\tau)=-\Psi_{-}^{i}(0,\tau),
\end{equation}
and
\begin{equation}
\epsilon_{+}=-\epsilon_{-}.
\end{equation}
However, having ruled out (ii) by our choice of boundary
conditions in the previous subsection, we are restricted to (i),
which eliminates half of the world-sheet supersymmetry.  We could
have reversed the fermion boundary conditions at the two ends and
consequently chosen (ii), but the obvious conclusion is that our
mixed Dirichlet-Neumann conditions are compatible with only half
of the usual supersymmetry transformations.  This breaking of
worldsheet SUSY will have to manifest itself in the space-time
spectrum, as we will see later.  Since Poincar\'{e} invariance has
been explicitly broken, we do not expect spacetime supersymmetry
operators, which are essentially square-roots of the Poincar\'{e}
generators, to be effective.

\subsection{Gauge Choice:}

For the bosonic string with one end fixed at the origin, it was
natural to choose the gauge condition

\begin{equation}
X^{0}=\alpha_{0}^{0}\tau =lp^0 \tau
\end{equation}
just as the light-cone gauge $X^{+}\sim\tau$ is suited to a string
with both ends moving at the speed of light.  Indeed, the first
excited state of such a free open string is a massless vector
boson with $D-2$ polarizations, which is the number of transverse
oscillators used in the light-cone gauge. Our problem, in
contrast, leads to a massive vector with $D-1$ polarization
states, which is in accord with (20). Combining this gauge choice
with (13), we get

\begin{equation} \delta X^{0}=\bar{\epsilon} \Psi^{0}=0.
\end{equation}

This is similar to setting all the $\Psi^{+}$ and $X^{+}$
oscillators to zero in the light-cone gauge treatment of the usual
RNS string.  The condition (21) kills all the $b_{n}^{0}$ modes of
the NS-sector.  On the other hand, for the R-sector, all
$d_{n}^{0}$ modes are zero except $d_{0}^{0}$. To see why this mode
survives, consider the explicit form of (21) with the $\Psi ^i$ in
(15) replaced with $\Psi ^0$ and $\epsilon_+ = \epsilon_-$.  This
is only compatible with $d_{0}^{0}$ being non-zero.  This is just
as well, since we would not want to lose an element of the
Clifford algebra in $(D-1,1)$.

\section{Quantization}

With the ``rest-frame gauge" chosen above,  the $\alpha_{n}^{0}$
modes and their associated negative norm  states \cite{Gursoy} are
discarded at the outset.  However, unlike the situation in the
light-cone gauge, this does not mean that all the states obtained
by hitting the ground state with the spacelike oscillators are
automatically physical. In the light-cone gauge, the constraints
are implemented at the operatorial level, allowing one to solve
for the $\alpha^{-}_n$ in terms of transverse oscillators.  Our
gauge choice does not involve an off-diagonal space-time
metric and does not offer a similar possibility.  There are still
constraints which must be imposed on all possible states, and only
the states annihilated by these constraints are the final physical
ones. This is similar to non-covariant treatments of gauge
theories where one first sets $A_{0}=0$, but then also imposes
$\nabla \cdot \textbf{A}=0$ to isolate the physical degrees of freedom.
Since negative metric states in the Hilbert space are eliminated 
from the beginning, Faddeev-Popov ghosts are not needed.

\subsection{Canonical Quantization:}

We will follow the conventional canonical quantization procedure.
The canonical anticommutation relations for the fermionic fields
are

\begin{equation}
\left\{\Psi_{\alpha}^{i}(\sigma, \tau), \Psi_{\beta}^{j}(\sigma',
\tau)\right\}=\pi\delta(\sigma-\sigma')\delta^{ij}\delta_{\alpha\beta}.
\end{equation}

Substituting the known mode expansions, we obtain anticommutation
relations for the modes.  These are

\begin{equation}
\left\{b_{r}^{i}, b_{s}^{j}\right\}=\delta^{ij}\delta_{r+s} (r, s:
odd )
\end{equation}
for the NS-sector, and

\begin{equation}
\left\{d_{m}^{i}, d_{n}^{j}\right\}=\delta^{ij}\delta_{m+n} (m,n:
even)
\end{equation}
for the R-sector. Finally, the $d_{0}^{0}$ mode obeys

\begin{equation}
\{d_{0}^{0}, d_{0}^{0}\}=-I.
\end{equation}

As in the ordinary RNS string, the $d_{0}^{i}$ and $d_{0}^{0}$ are
proportional to the Dirac gamma matrices with $\Gamma^{\mu} = i
\sqrt{2} d^{\mu}_0 $ in $(D-1,1)$ dimensional Minkowski space.

\subsection{Super-Virasoro constraints}

The unphysical modes are eliminated by imposing
$T^{+}_{\alpha\beta}\mid\Psi_{phys}\rangle=0$ and
$J^{+}_{\alpha}\mid\Psi_{phys}\rangle=0$ on the states, where
$T^{+}_{\alpha\beta}$ and $J^{+}_{\alpha}$ are the
positive-frequency components of the energy-momentum tensor and
the supercurrent, respectively.

The energy-momentum tensor for the fermionic part is
\begin{equation}
T_{\alpha\beta}=\frac{i}{4}\bar{\Psi}^{\mu}\rho_{\alpha}\partial_{\beta}\Psi_{\mu}+
\frac{i}{4}\bar{\Psi}^{\mu}\rho_{\beta}\partial_{\alpha}\Psi_{\mu}-
\frac{i}{4}\eta_{\alpha \beta}\bar{\Psi}^{\mu}\rho^{\gamma}\partial_{\gamma}\Psi_{\mu}
\end{equation}
and, in the chosen gauge, its components in conformal coordinates
are
\begin{eqnarray}
T_{++}=\frac{1}{2}(T_{00}+T_{01})&=&\frac{i}{2}:\Psi_{+}^{i}\partial_{+}\Psi_{+i}:,\\
T_{--}=\frac{1}{2}(T_{00}-T_{01})&=&\frac{i}{2}:\Psi_{-}^{i}\partial_{-}\Psi_{-i}:.
\end{eqnarray}
The supercurrent is
\begin{equation}
J_{\alpha}=\frac{1}{2}\rho^{\beta}\rho_{\alpha}\Psi^{\mu}\partial_{\beta}X_{\mu}.
\end{equation}
The nonzero components of the supercurrent, again in conformal
coordinates, are
\begin{eqnarray}
J_{+}&=&\Psi_{+}^{i}\partial_{+}X_{i}+\Psi_{+}^{0}\partial_{+}X_{0},\\
J_{-}&=&\Psi_{-}^{i}\partial_{-}X_{i}+\Psi_{-}^{0}\partial_{-}X_{0}.
\end{eqnarray}
The conservation law of the supercurrent yields
\begin{equation}
\partial_{-}J_{+}=\partial_{+}J_{-}=0.
\end{equation}
We now examine the implications of these constraints in the
Neveu-Schwarz and Ramond sectors.

\subsubsection{Neveu-Schwarz constraints}

The super-Virasoro operators are the Fourier coefficients of the
energy-momentum tensor and supercurrent. To obtain these
coefficients we use the fact that $T_{++}(\sigma)=T_{--}(-\sigma)$
and the ``doubling trick" of Polchinski \cite{Pol}, extending the
region of integration to $[-\frac{\pi}{2},\frac{\pi}{2}]$.  We can
then write
\begin{equation}
L_{m}^{(b)}=\frac{1}{\pi}\int_{\frac{-\pi}{2}}^{\frac{\pi}{2}}e^{im\sigma}
T_{++}(\sigma)d\sigma.
\end{equation}
Strictly speaking, for $m=0$ the above definition disagrees by an 
additive constant with what 
should properly be called $L_{0}$.  The constant 
arises from the non-tensor Schwarzian derivative term in the 
transformation 
from the $(\tau,\sigma)$ coordinates used above to the $z$ coordinate 
($z=exp(-i\sigma +\tau)$ in the Euclideanized world sheet) in terms of 
which the $L_{n}$ are actually defined. We will later compute such 
constants. Substituting the mode expansions, we get

\begin{equation}
L_{m}^{(b)}=\frac{1}{2}\sum_{r:odd}(r+\frac{m}{2}):b_{-r}^{i}b_{m+r}^{i}:+a^{b}\delta_{m,0}
\end{equation}
where $m$ is seen to be necessarily even. It is  normal ordering for $m=0$ that brings the constant
$a^{b}$.  The regularization recipe for calculating such 
constants for all modes will be given later.  The $L_{m}^{(b)}$ thus
complement the $L_{m}^{(\alpha)}$ which are also even.  This is an
indication of the consistency of the mode elimination scheme
imposed by our mixed boundary conditions.

The fermionic generators of this sector are obtained similarly
using $J_{+}(\sigma)=J_{-}(-\sigma)$, which results in

\begin{equation}
G_{m}=\frac{\sqrt{2}}{\pi}\int_{\frac{-\pi}{2}}^{\frac{\pi}{2}}e^{im\sigma}J_{+}(\sigma)d\sigma.
\end{equation}
In terms of the mode expansions one finds
\begin{equation}
G_{m}=\sum_{r:odd}\alpha_{-r}^{i}b_{m+r}^{i}
\end{equation}
where $m$ is again forced to be even.

The super-Virasoro algebra for this sector is

\begin{eqnarray}
\left[L_{m},L_{n}\right]&=&(m-n)L_{m+n}+m(\alpha^{0}_{0})^{2}\delta_{m+n} 
+ A(m)\delta_{m+n},\\
\left[L_{m},G_{n}\right]&=&(\frac{m}{2}-n)G_{m+n},\\
\left\{G_{m},G_{n}\right\}&=&2L_{m+n}-2(a^{\alpha} 
+a^{b})\delta_{m+n}+(\alpha^{0}_{0})^{2}\delta_{m+n} + B(m)\delta_{m+n}
\end{eqnarray}
where $L_{m}=L_{m}^{(\alpha)}+L_{m}^{(b)}$ and

\begin{eqnarray}
L_{0}^{(\alpha)}&=&\frac{-1}{2}(\alpha_{0}^{0})^{2}+\frac{1}{2}\sum_{r:odd}:\alpha_{r}^{i}\alpha_{-r}^{i}:
+a^{\alpha}, \\
L_{m}^{(\alpha)}&=&\frac{1}{2}\sum_{r:odd}:\alpha_{r}^{i}\alpha_{m-r}^{i}:,
\end{eqnarray}
where, in the second line, m is even and nonzero.  The normal ordering constant $a^{\alpha}$ will be 
presented later.

We see that all super-Virasoro operators have their usual forms,
say as in \cite{GSW}, except for the fact that they are written in
terms of odd modes only. We must now see how this changes the central
extension terms. The $\alpha$-mode anomaly $A(m)$ was
calculated in \cite{Gursoy} with the result

\begin{equation}
A^{\alpha}(m)=\frac{(D-1)}{24}(m^{3}+2m).
\end{equation}

Thus we will only calculate the anomaly due to the $b$-modes, and
this can be easily done via

\begin{eqnarray}
A^{(b)}(m)&=&\langle0\mid[L_{m}^{(b)},L_{-m}^{(b)}]\mid0\rangle\\
          &=&\langle0\mid L_{m}^{(b)}L_{-m}^{(b)}\mid0\rangle\\
          &=&\frac{1}{4}\langle0\mid\sum_{r,s:odd}(r+\frac{m}{2})(s-\frac{m}{2})b_{-r}^{i}b_{m+r}^{i}b_{-s}^{j}b_{-m+s}^{j}\mid0\rangle\\
          &=&\frac{1}{4}\langle0\mid\sum_{r,s:odd}(r+\frac{m}{2})(s-\frac{m}{2})b_{m+r}^{i}b_{-r}^{i}b_{-m+s}^{j}b_{-s}^{j}\mid0\rangle,
\end{eqnarray}
where m is, of course, even and positive.  The expectation value
in the third line vanishes for $r>0$ and $s>m$, whereas the
expectation value in the fourth line vanishes for $s<0$ and
$m<-r$.  This yields

\begin{eqnarray}
A^{(b)}(m)&=&\frac{1}{4}\langle0\mid\sum_{r=-m+1}^{-1}\sum_{s=1}^{m-1}(r+\frac{m}{2})(s-\frac{m}{2})b_{-r}^{i}b_{m+r}^{i}b_{-s}^{j}b_{-m+s}^{j}\mid0\rangle, \\
          &=&2(D-1)\sum_{s=1}^{m-1}(s-\frac{m}{2})^{2}
\end{eqnarray}
where all the summations above are over odd integers and
$\eta_{ij}\eta^{ij}=D-1$ is the number of space coordinates.  The
summation in the second line, when taken over all integers,
produces the well-known NS-sector central extension term.  On the
other hand, our odd integer summation gives

\begin{equation}
A^{(b)}(m)=\frac{(D-1)}{48}(m^{3}-4m).
\end{equation}

Hence the total central extension term of the first commutator is
\begin{equation}
A(m)=\frac{(D-1)}{16}m^{3}.
\end{equation}

A similar computation for the
central extension term appearing in the anti-commutator of the
$G_m$ gives
\begin{eqnarray}
B(m)&=&\langle0\mid G_{m}G_{-m}\mid0\rangle, \\
    &=&\frac{(D-1)}{4}m^{2}.
\end{eqnarray}

At this point, we note that the anomaly-free $OSp(1\mid2)$ closed
subalgebra of the NS-sector of the usual RNS string is now reduced
to the Abelian super-subalgebra of $L_{0}$ and $G_{0}$.

The mass-shell condition comes from the zero-frequency part of the
Virasoro constraint $T_{\alpha\beta}=0$.  At the quantum level,
the masses of the NS states follow from

\begin{equation}
(L_{0}^{(\alpha)}+L_{0}^{(b)})\mid\Psi_{phys}\rangle=0
\end{equation}

\subsubsection{Ramond constraints}

The super-Virasoro operators of this sector  are

\begin{equation}
L_{m}^{(d)}=\frac{1}{\pi}\int_{\frac{-\pi}{2}}^{\frac{\pi}{2}}e^{im\sigma}T_{++}(\sigma)d\sigma
\end{equation}
and

\begin{equation}
F_{m}=\frac{\sqrt{2}}{\pi}\int_{\frac{-\pi}{2}}^{\frac{\pi}{2}}e^{im\sigma}J_{+}(\sigma)d\sigma.
\end{equation}
Substituting the known mode expansions of $T_{++}$ and $J_{+}$, we
find

\begin{equation}
L_{m}^{(d)}=\frac{1}{2}\sum_{n:even}(n+\frac{m}{2}):d_{-n}^{i}d_{m+n}^{i}:+a^{d}\delta_{m,0}
\end{equation}
where m must be even, and

\begin{equation}
F_{s}=\sum_{r:odd}\alpha_{-r}^{i}d_{r+s}^{i}
\end{equation}
where s is  necessarily odd.  An exception is the operator

\begin{equation}
f_{0}=\alpha_{0}^{0}d_{0}^{0}
\end{equation}
which is all that survives here from  the generalized Dirac
operator $F_0$ of the standard RNS string. The super-Virasoro algebra of
this sector is as follows:

\begin{eqnarray}
\left[L_{m},L_{n}\right]&=&(m-n)L_{m+n}+A(m)\delta_{m+n}+ 
m (\alpha^{0}_{0})^{2} \delta_{m+n},\\
\left[L_{m},F_{r}\right]&=&(\frac{m}{2}-r)F_{m+r},\\
\left\{F_{r},F_{s}\right\}&=&2L_{r+s} 
+(\alpha^{0}_{0})^2
\delta_{r+s}-2(a^{\alpha}+a^{d})\delta_{n+m} +
B(r)\delta_{r+s},\\
\left[L_{m}, f_{0}\right]&=&0, \\
\left\{f_{0}, f_{0}\right\}&=&-(\alpha_{0}^{0})^{2}
\end{eqnarray}
where $L_{m}=L_{m}^{(\alpha)}+L_{m}^{(d)}$ and the commutators (or
anticommutators, as appropriate) of $(\alpha_{0}^{0})^{2}$ or
$(f_{0})$ with other operators vanish.

The central extension terms of this algebra can be calculated as
in the previous section, giving

\begin{equation}
A^{(d)}(m)=\frac{(D-1)}{48}(m^{3} +8m).
\end{equation}

Then the total central extension term of the first commutator,
including the contribution of the $\alpha$ modes becomes

\begin{equation}
A(m)=\frac{(D-1)}{16}(m^{3}+4m).
\end{equation}

A similar treatment for the central extension term of
anti-commutators of fermionic generators yields

\begin{equation}
B(r)=\frac{(D-1)}{4}(r^{2}+1).
\end{equation}

\section{The Spectrum}

\subsection{Normal ordering constants:}

In this section we will compute the normal ordering constants $a^{\alpha}, 
a^{b}$ and $a^{d}$ using Polchinski's `heuristic recipe' \cite{Polrecipe}.  
This involves two steps: 1) One adds the oscillator zero point energies, 
regularizing the divergent result via an analytic continuation of 
the $\zeta$-function. This gives the normal ordering constant in the 
$(\tau,\sigma)$ coordinates. 2) The $L_{n}$, and in particular $L_{0}$ are 
by contrast the modes in the expansion of $T(z)$ in powers of the $z$ 
coordinate (on 
the Euclideanized world sheet, $z=\exp (-i \tau + \sigma)$); thus one 
must add the non-tensor shift coming from the Schwarzian derivative of the 
transformation from $(\tau,\sigma)$ to $z$.   For $L_{0}$, the shift is 
given 
by $c \frac{1}{24}$, where $c$ is the central charge. In \cite{Gursoy}, 
this constant contribution was omitted; the masses there should all be 
increased by $\frac{D-1}{48}$.    

In our case, the total $c$ 
is $\frac{D-1}{2} (1+ \frac{1}{2})$, where $D-1$ obviously 
comes from the number of space coordinates, while the $2$ in the 
denominator 
can be interpreted as the result of keeping of only odd or even modes. 
The last $3/2$ expresses the fact that each world-sheet boson contributes 
$1$ and each world-sheet 
fermion $1/2$. The combined shift of $\frac{D-1}{32}$ thus applies to 
both the R and the NS sectors.  

We may similarly compute the normal ordering constants for each 
excitation seperately. 
The zero point energy of the $\alpha$ oscillators is 
$\frac{D-1}{2}$ times
a sum over positive odd integers, which, using $\zeta(-1)=\frac{-1}{12}$, 
is simply $\zeta(-1)-2\zeta(-1)=\frac{1}{12}$.  The shift is 
$\frac{D-1}{48}$, yielding $a^{\alpha}=\frac{D-1}{16}$.
The zero point energy of the odd integral moded NS $b$ world-sheet 
fermion oscillators is $ - \frac{D-1}{2}\frac{1}{12}$ and the shift is 
$\frac{D-1}{96}$, giving $a^{b}= - \frac{D-1}{32}$. The total normal 
ordering constant in $L_{0}(NS)$ is thus $\frac{D-1}{32}$.
For the $d$ oscillators, the zero point energy is 
$\frac{D-1}{2}(-\frac{1}{2})(-\frac{1}{6})$, the first minus being due to 
the fermionic nature of the $d$'s and the $(-\frac{1}{6})$ coming from a 
sum of even positive integers.  With the shift of $\frac{D-1}{96}$, we 
find $a^{d}=  \frac{3(D-1)}{32}$ and the total normal
ordering constant in $L_{0}(R)$ becomes $ \frac{5(D-1)}{32}$.

\subsection{Mass formulas in the NS- and R-Sectors:}

With the calculated normal ordering constants, we can write down
the mass-shell conditions.  For the NS-sector, we have
\begin{equation}
(-\alpha'_{ND}M^{2}+N^{(\alpha)}+N^{(b)}+\frac{D-1}{32})\mid\Psi_{phys}\rangle=0,
\end{equation}
while for the R-sector
\begin{equation}
(-\alpha'_{ND}M^{2}+N^{(\alpha)}+N^{(d)}+
\frac{5(D-1)}{32})\mid\Psi_{phys}\rangle=0,
\end{equation}
where $\alpha'_{ND}=\frac{l^{2}}{2}$, as calculated in \cite{Gursoy}.
This doubling of the slope relative to $\alpha'_{NN}$ of the usual
open string is one of the principal distinguishing features of our
system. A quick way to understand this result is to consider a
classical string with one end fixed in its highest angular
momentum (leading Regge trajectory) mode, where it rotates
rigidly. This can also be viewed as an ordinary open string in the
same mode with its center of mass at rest and then throwing away
one half.  In order to preserve the  relation $J \sim \alpha' M^2$
with $J$ and $M$ both being halved, the slope has to be doubled.

Since half of the world-sheet supersymmetry has been broken at the
beginning by the restriction $(\epsilon_{+}=\epsilon_{-})$, we do
not expect  space-time supersymmetry to appear the way it does in
the RNS string subjected to the GSO projection \cite{GSO}. Actually, the
halving of supersymmetry via $D$-branes is a familiar
phenomenon \cite{Pol2}, and the fixed end here is nothing but a
$D0$-brane.  While the spectrum is thus not expected to be fully
supersymmetric, we have two further options in the degree of
superymmetry breaking:

(i) $D-1=8n$:   The masses of the states in the NS-sector have the
the same values as those in the R-sector for $\alpha'_{ND}M^{2}\geq
n$, partially preserving SUSY in the mass spectrum.

(ii) $D-1 \neq 8n$:   Mass values in the two sectors are
completely different and SUSY is completely broken.

In the following, we will work with the minimal supersymmetry breaking
option $n=1$, $D=9$. All states being
massive, we expect to see irreps of $SO(8)$, which is clearly a
remnant of the space-time SUSY enjoyed by the ten-dimensional
superstring. We now examine the spectra of the two sectors
separately.

\subsection{Neveu-Schwarz spectrum:}
A physical state in the NS-sector must satisfy

\begin{eqnarray}
G_{2m}\mid\phi\rangle&=&0,   m>0\\
L_{2n}\mid\phi\rangle&=&0,   n>0\\
L_{0}\mid\phi\rangle&=&0,
\end{eqnarray}
and it is the last condition that leads to
$\alpha'_{ND}M^{2}=N^{(\alpha)}+N^{(b)}+\frac{D-1}{32}$.

There is also a  $G_0$ constraint that has to be handled
separately.  An examination of the NS constraint algebra (37-39)
shows that all the physical state conditions can be obtained from
$G_2$, $G_0$ and $L_0$. $G_0$ does not in fact annihilate
physical states; it instead requires them to be its eigenstates with
the masses as eigenvalues.  From (39) and (67), we see that

\begin{equation}
G_0 \mid\phi\rangle=\sqrt{ N^{(\alpha)}+N^{(b)} }M\mid\phi\rangle
\end{equation}
We note that this amounts to taking the square root of the
Klein-Gordon-like operator $L_0$.  The novelty is that this
happens not in the fermionic but in the bosonic sector!

We now examine how low-lying physical states are obtained.  Since
negative-metric states have been barred from the beginning, it is
not immediately obvious what role is left for the above
constraints to play.  If we use the $L_n$ directly, the answer in
the bosonic NS sector turns out to be that all $\alpha_{-n}^{i}$
and $b_{-n}^{i}$ oscillators for $n>1$ are ruled out, and the
surviving states are automaticaly organized into $SO(8)$ irreps.
Hence the daughter trajectories are eliminated from the spectrum.
The $G_n$ constraints prune the remaining states even further; for
example, a potential $N=3$ state of the form
$\alpha_{-1}^{i}\alpha_{-1}^{j}b_{-1}^{k}\mid0\rangle$ is
prohibited by the $G_{2}$ constraint. Finally, the $G_0$
constraint allows only specific linear combinations of the states
that have survived that far. This is obviously different from what
happens in the light-cone gauge in the usual RNS string, where all
combinations of $\alpha_{-n}^{i},b_{-n}^{i}, (i=1,...,8)$
oscillators on the vacuum are guaranteed to produce physical
states, which then combine with the others at the same mass to
give $SO(9)$ irreps.  There being no obvious pattern to the
allowed states beyond what we have just mentioned, we limit
ourselves to displaying below the contents of the first four
levels.

\begin{itemize}
\item $N=0, M^{2}=0$
\begin{center}
$\mid0\rangle$
\end{center}
ia a massless scalar, providing  a stable vacuum for this sector.
\item $N=1$, $\alpha'_{ND}M^{2}=1$: We start with the two massive
vector states
\begin{center}
$\mid\alpha\rangle \equiv  \alpha_{-1}^{i}\mid0\rangle, {\bf 8_{v}}$\\
\vspace{0.2cm}
 $\mid\beta\rangle \equiv  b_{-1}^{i}\mid0\rangle,{\bf 8_{v}} $
\end{center}
allowed by the $G_2$ constraint.  However, the only combination
permitted by $G_0$ is $\mid\alpha\rangle + \mid\beta\rangle $.
\item $N=2$, $\alpha'_{ND}M^{2}=2$: Under $G_2$, the massive
tensor states
\begin{eqnarray*}
\mid1\rangle &\equiv&
\left\{\alpha_{-1}^{i}\alpha_{-1}^{j}-\frac{\delta^{ij}}{D-1}\alpha_{-1}^{k}\alpha_{-1}^{k}\right\}\mid0\rangle, {\bf 35_{v}}\\
 \mid2\rangle &\equiv&
b_{-1}^{i}b_{-1}^{j}\mid0\rangle, {\bf
28}\\
\mid3\rangle &\equiv&
\left\{\alpha_{-1}^{i}b_{-1}^{j}+\alpha_{-1}^{j}b_{-1}^{i}-\frac{2\delta^{ij}}{D-1}\alpha_{-1}^{k}b_{-1}^{k}\right\}\mid0\rangle, {\bf 35_{v}}\\
\mid4\rangle &\equiv&
\left\{\alpha_{-1}^{i}b_{-1}^{j}-\alpha_{-1}^{j}b_{-1}^{i}\right\}
\mid0\rangle,{\bf28}
\end{eqnarray*}
are the allowed combinations.  We recall that since the even
$\alpha$ and b modes have been eliminated at the beginning, we
cannot have excitations such as $\alpha_{-2}^{i}\mid0\rangle$ and
$b_{-2}^{i}\mid0\rangle$. Out of the above four, the condition
$G_0\mid\phi\rangle =\sqrt{2}\mid\phi\rangle$ allows only
\[
\sqrt 2\mid1\rangle + \mid3\rangle,  {\bf35_{v}}
\]
and
\[
\sqrt 2\mid2\rangle + \mid4\rangle, {\bf28_{v}}.
\]
\item $N=3$, $\alpha'_{ND}M^{2}=3$
\begin{eqnarray*}
\mid1\rangle &\equiv&
\{\alpha_{-1}^{i}\alpha_{-1}^{j}\alpha_{-1}^{k}-\frac{\delta^{ij}}{D+1}\alpha_{-1}^{n}\alpha_{-1}^{n}\alpha_{-1}^{k}-\frac{\delta^{ik}}{D+1}\alpha_{-1}^{n}\alpha_{-1}^{n}\alpha_{-1}^{j}\\
&\phantom{-}&\phantom{\{\alpha_{-1}^{i}\alpha_{-1}^{j}\alpha_{-1}^{k}-\frac{\delta^{ij}}{D+1}\alpha_{-1}^{n}\alpha_{-1}^{n}\alpha_{-1}^{k}}-\frac{\delta^{jk}}{D+1}\alpha_{-1}^{n}\alpha_{-1}^{n}\alpha_{-1}^{i}\}\mid0\rangle,{\bf112_{v}}\\
 \mid2\rangle &\equiv&
\{b_{-1}^{i}b_{-1}^{j}\alpha_{-1}^{k}+b_{-1}^{k}b_{-1}^{j}\alpha_{-1}^{i}+b_{-1}^{i}b_{-1}^{k}\alpha_{-1}^{j}+\frac{2\delta^{jk}}{2-D}b_{-1}^{i}b_{-1}^{n}\alpha_{-1}^{n}\}\\
&\phantom{-}&\phantom{\{b_{-1}^{i}b_{-1}^{j}\alpha_{-1}^{k}+b_{-1}^{k}b_{-1}^{j}\alpha_{-1}^{i}+b_{-1}^{i}b_{-1}^{k}\alpha_{-1}^{j}}-\frac{2\delta^{ik}}{2-D}b_{-1}^{j}b_{-1}^{n}\alpha_{-1}^{n}\}\mid0\rangle, {\bf 160_{v}}\\
\end{eqnarray*}
\begin{eqnarray*}
\mid3\rangle &\equiv&
\{b_{-1}^{i}b_{-1}^{j}b_{-1}^{k}\}\mid0\rangle, {\bf 56_{v}}\\
\mid4\rangle &\equiv&
\{b_{-1}^{i}b_{-1}^{j}\alpha_{-1}^{k}-b_{-1}^{i}b_{-1}^{k}\alpha_{-1}^{j}-b_{-1}^{k}b_{-1}^{j}\alpha_{-1}^{i}\}\mid0\rangle,{\bf56_{v}}
\end{eqnarray*}
\end{itemize}

The tensorial form of the first state is dictated by $L_{2}$,
which is obtained from the anticommutator of $G_2$ and $G_0$. As
mentioned earlier, the same $L_2$ prohibits states with higher
oscillators of the form $\alpha_{-3}^{i}\mid0\rangle$, The forms
of the second and fourth states are determined by the action of
$G_{2}$. Finally, only the combination
\[
\sqrt{3}\mid3\rangle + \mid4\rangle,  {\bf 56_{v}}
\]
obeys $G_0\mid\phi\rangle = \sqrt{3}\mid\phi\rangle $.

\subsection{Ramond spectrum:}

The physical states in the Ramond sector must satisfy

\begin{eqnarray}
F_{2m+1}\mid\psi\rangle&=&0,   m>0\\
L_{2n}\mid\psi\rangle&=&0,   n>0\\
(L_{0})\mid\psi\rangle&=&0.
\end{eqnarray}
However, using the superalgebra (59-63), one can see that these
infinite set of conditions can be reduced to just the $F_1$, $ F_3$
and $L_0$ constraints.  In addition, taking the square root of
(63), we have an $f_0$ constraint which simplifies to
\begin{equation}
(\Gamma^0 + I)\mid\psi\rangle = 0.
\end{equation}

This is what remains of the Dirac equation.  We must now discuss
the properties of the Ramond ground state $\mid\psi\rangle$ and
the meaning of (76).

Majorana spinors with 16 components (and depending on 16 real
parameters) are allowed in our $(8,1)$ Minkowski space-time.
These 16 components consist of linear combinations of the two
independent $SO(8)$ spinors, which we will denote by ${\bf 8_{s}}$
and ${\bf 8_{c}}$.  These are projected out of a 16-component
spinor by the operators $(\Gamma^0 + I)$ and $(\Gamma^0 - I)$,
respectively.  Thus (76) indeed serves as a Dirac equation in
halving the number of independent components.

We now apply combinations of creation operators with $N=0,1,2$ on
ground states $\mid\psi_0\rangle$, and then obtain the physically
allowed combinations by imposing the $F_{1}, F_{3},  L_0$ and $f_0$
conditions:

\begin{itemize}
\item $N=0, \alpha'_{ND}M^{2}=1$
\[
\mid0\rangle \equiv \mid\psi_{\beta}\rangle, \psi_{\beta} \sim
{\bf 8_{s}}
\]
Note the $f_0$ constraint (76) has eliminated ${\bf 8_{c}}$ and kept
${\bf 8_{s}}$.
\item $N=1, \alpha'_{ND}M^{2}=2$
\[
\mid1\rangle \equiv \mid\psi^{ij}_{\beta'}\rangle \equiv
(\alpha_{-1}^{i}d_{0}^{j}+\alpha_{-1}^{j}d_{0}^{i}-
\frac{\delta^{ij}}{D-1}\alpha_{-1}^{k}d_{0}^{k})
\mid\psi_{\beta'}\rangle \sim {\bf 35_{v}\otimes \bf 8_{c}}
\]
is the only permissible combination.
Imposing $(\Gamma^0 + I)\mid1\rangle = 0$
forces $\mid\psi_{\beta'}\rangle$ to be ${\bf 8_{c}}$  since $\Gamma^0$
anticommutes with $d^{i}_{0}$.
\item $N=2,\alpha'_{ND}M^{2}=3$
\begin{eqnarray*}
\mid\psi^{ij}_{\beta}(\alpha \alpha)\rangle
+\mid\psi^{ij}_{\beta}(dd)\rangle &\equiv&
(\alpha_{-1}^{i}\alpha_{-1}^{j}-\frac{\delta^{ij}}
{D-1}\alpha_{-1}^{k}\alpha_{-1}^{k}\\&-&d_{-2}^{i}d_{0}^{j}-
d_{-2}^{j}d_{0}^{i}+\frac{2\delta^{ij}}{D-1}d_{-2}^{k}d_{0}^{k})
\mid\psi_{\beta}\rangle,\sim {\bf 35_{v}\otimes\bf 8_{s}}
\end{eqnarray*}
\end{itemize}

The $\alpha \alpha$ and $dd$ parts of this separately satisfy the
Virasoro constraints (74) and (75), but the superconstraints (73) force
them into this particular combination. Because of the even number
of $d$ modes, the basic spinor is ${\bf 8_{s}}$.

A clarification about these states is in order. The boldface numbers refer 
only to  the $SO(8)$ transformation properties of 
the states above.  This is because of the fact that the R-sector $SO(8)$ 
generators are built out of 
$\alpha^{i}_{n}$'s and $d^{i}_{n}$'s, and thus transform the R-states 
exactly as indicated as in boldface. 
However, one must be careful in distinguishing between these 
numbers and the actual physical degrees of freedom.  The
NS spectrum, where the boldface numbers are identical with the
number of physical states, is free of this complication.
We must now count the true number of physical states in the R sector.

Staring with the $N=1$ state, we see  that we cannot be
dealing with 35x8 physical states: the $d^{i}_{0}$ merely shuffle the 8
components of the ground state spinor.  Thus we have at most 8x8 = 64
states, but since the tensor is traceless, 8 spinor components
corresponding to $\psi^{ii}_{\beta'}$ are absent, and the true physical
content is ${\bf {56}_{s}}$.  This is not as unfamiliar a situation as it
might first appear: consider a scalar field $\theta$ and its gradient
$\partial_{\mu}\theta$.  The latter transforms as a $D$-vector, but the
physical degree of freedom is still just $\theta$, which is a $D$-scalar.

Examining the physical content of the $N=2$ states, we see that
$\mid\psi^{ij}_{\beta}(\alpha \alpha)\rangle$ contains 35x8 = 280 states,
but these do not belong to a single irrep.  Among the 280 states there is
a ${\bf {56}_c}$ of the form $d^{i}_{0}\mid\psi^{ij}_{\beta}(\alpha
\alpha)\rangle$ and the rest is ${\bf {224}_c}$.  The part
$\mid\psi^{ij}_{\beta}(dd)\rangle $ represents another ${\bf {56}_c}$,
because the $d^{i}_{0}$ does not increase the number of states
(bringing the number down from 280 to 64) and tracelessness in $ij$
takes off another 8.  Thus the $N=2$ content is ${\bf {224}_c}+{\bf
{56}_c} + {\bf {56}_c}$.  Counting the $\alpha$ and $d$ modes separately
and adding the numbers may seem surprising, but it is again not new: in
the Higgs phenomenon, the massless photon field $A_{\mu}$ and
$\partial_{\mu}\theta$ are added to form the massive vector
field $B_{\mu}$.  Although all three formally transform as 4-vectors,
the final $B_{\mu}$ has 2+1 = 3 degrees of freedom.

The GSO projection in its original form turns out to be inapplicable to the mixed boundary
condition superstring.  Recall that the GSO operation projects out 
one of the two chiralities of fermionic states; this projection is impossible in
principle in our odd-dimensional spacetime ( in the R-sector we have
the remnant Dirac equation operators $\Gamma^0 \pm I$, but not in the role
they play in the GSO projection).  A consequence of the the absence of the GSO procedure 
is the fact that low-lying states do
not exhibit space-time supersymmetry beyond matching masses and Regge slopes
in the two sectors, but there are more basic manifestations of the
inapplicability.  For example, our final physical states are not
homogeneous in the number of
space-time fermions.  Since the mode structure here is
different from the standard RNS string (for example, our $b$ operators
add an integer rather than half  an integer to the squared mass), one
could not in any case have expected the GSO procedure to work in the usual
way.

\subsection{Density of states:}

Upper bounds for the number of bosonic and fermionic excitations at a given level can be
calculated by slightly modifying  standard techniques. The most important 
difference between the standard light-cone
calculation and the one here is that our calculation gives the number of states before
Virasoro and super-Virasoro constraints are applied. Thus the total number of pre-constraint 
string states at each level is given by $tr\omega^{N}$.  Now let us calculate this trace for the
two sectors separately.
 
(i) Neveu-Schwarz Sector:
 
Counting only odd $\alpha$ and $b$ modes leads to the result
\begin{eqnarray}
g_{NS}(\omega)&=&tr\omega^{N}\nonumber\\
&=&tr\omega^{N_{\alpha}+N_{b}}\nonumber\\
&=&\{\prod_{r=1,3,5 ...}\frac{1+\omega^{r}}{1-\omega^{r}}\}^{(D-1)}.
\end{eqnarray}
Note that the GSO projection operator found in the usual light-cone 
formula is absent here.

(ii) Ramond Sector:
 
Even moded $d$ and odd moded $\alpha$ oscillators give
\begin{eqnarray}
g_{R}(\omega)&=&\lambda \times tr\omega^{N}\nonumber\\
&=&\lambda \times tr\omega^{N_{\alpha}+N_{d}}\nonumber\\
&=&\lambda \times \omega^{\frac{D-1}{8}}\{\prod_{r=1,3,5
...}\frac{1+\omega^{r+1}}{1-\omega^{r}}\}^{(D-1)}.
\end{eqnarray}
The factor $\lambda$ represents the degeneracy of the spinor ground state.  Prior to the $f_0$
constraint, $\lambda = 16$.  $\omega^{\frac{D-1}{8}}$ is due 
to the fact that the
ground state of
this sector has $\frac{D-1}{8}$ unit higher mass than that of the
Neveu-Schwarz sector.

For $D=9$ the expressions have the following expansions:
\begin{eqnarray}
g_{NS}(\omega)&=&1+16\omega+128\omega^2+704\omega^3+3072\omega
^4+11488\omega^5+38400\omega^6+117632\omega^7+... ,\nonumber\\
g_{R}(\omega)&=&16\omega+128\omega^2+704\omega^3+3072\omega^4+11488\omega^5+38400\omega^6+117632\omega^7...\nonumber.
\end{eqnarray}

We have unfortunately not been able to find an analog of Jacobi's 
``abstruse identity" showing $g_{NS}(\omega)-1 = g_{R}(\omega)$ to all 
orders.  Assuming the multiplicities continue to be equal, the surprising 
conclusion seems to be that the halving of 
world-sheet supersymmetry shows its effects on the spacetime spectumm not 
directly, but through the (super)constraints.

\section{Discussion}

Our concern in this note has been twofold: The first was to see whether an
RNS string with one fixed end would provide qualitative
phenomenological hints and distinctive signatures about the spectrum of
hadrons with one very heavy quark.  The second concern was the more formal
one of working out the novel consequences of the unusual mixed
Dirichlet-Neumann boundary conditions and seeing  whether the
resulting system can be quantized in a consistent way. Just as in the
original attempts to apply string theory to hadrons, these two aims are
partially in conflict, most notably in the dimension of space-time.

Ignoring this conflict as was customarily done in the old string-based
hadrodynamics, we can summarize some basic features of our model that may 
actually be observed in hadrons with one very heavy quark as follows:
(i) There is no
space-time supersymmetry except for the fact that meson and baryon Regge
trajectories have the same slope, leading to equal meson and baryon masses
for the higher states. (ii)  The Regge slope is twice that of the one
observed for lighter hadrons. 

An interesting point here is that if
we consider an interaction where the free
ends of two of our strings  join  while the heavy quark ends are kept
fixed, we are led to a model in which $X^{\mu}\sim \Sigma
\alpha_{n}^{\mu}e^{-in{\tau}}sin(n\sigma)$ with $\sigma$ in the range
$[0,\pi]$.  For such strings with both ends fixed on $D0$-branes, all
$\alpha_n$ modes and, consequently, all $L_n$ are allowed.  In this
sector, there will be ``daughter excitations", with the same spin but
equally spaced masses. but not in the sense there
are any leading linear Regge trajectories above them (the latter cannot be
present  because the
balancing of centrifugal force against tension in a rigid rotation
mode that defines leading trajectories is imposible when both
ends are fixed). This is in qualitative, and one hopes, not entirely
fortuitous  accord with the multiplicity of
$b\overline{b}$ and $c\overline{c}$ states of spin one. In any case,
the Coulomb part of the QCD potential is known to play an
important role in the dynamics of heavy quark-antiquark systems, ensuring
deviations from a mass spectrum based on the purely string-based picture.

Turning to string-theoretic issues, we start with the question of whether
using the more conventional $\sigma$ range $[0,\pi]$ would have made a
physical difference. One may  anticipate that the simultaneous doubling of
the $\sigma$ range and the
halving of the mode index $n$ will result in an equivalent physical
system, and this is indeed what happens.  The mixed boundary conditions now 
produce half-integral $\alpha$ and $b$ modes and integral Ramond $d$ 
modes, but
the allowed physical spectrum is exactly as the one above except for an
overall scaling.  In particular,
the preference for a $(8,1)$ space-time and $SO(8)$ symmetry remain.
We prefer working with odd $\alpha$ and $b$ and even $d$ modes because this
leads to the disappearence of odd $L_n$'s.  We have broken
Poincar\'{e} invariance by fixing one end of the string at a
special point in space; this is consistent with discarding the
operators  $L_1$ and $L_{-1}$ which involve the generator of space
translations.

\subsection*{Acknowledgements}

We are indebted to M. Ar\i k, I. Bars, S. De\~{g}er, C. Deliduman, M. 
Duff,
R. G\"{u}ven, A. Kaya, E. Sezgin, P. Sundell and G. Thompson for helpful
discussions. \

\end{document}